# Graph theory enables drug repurposing

# How a mathematical model can drive the discovery of hidden Mechanisms of Action

Ruggero Gramatica[a], T. Di Matteo[a], Stefano Giorgetti[b], Massimo Barbiani[b], Dorian Bevec[c], Tomaso Aste[b]

King's College London, Strand, London WC2R 2LS, United Kingdom – b. Department of Computer Science, University College London, Gower Street, London WC1 E6BT. c. Charité - Universitätsmedizin Berlin, Charitéplatz 1, 10117 Berlin

**ABSTRACT**

**We introduced a methodology to efficiently exploit natural-language expressed biomedical knowledge for repurposing existing drugs towards diseases for which they were not initially intended. Leveraging on developments in Computational Linguistics and Graph Theory, a methodology is defined to build a graph representation of knowledge, which is automatically analysed to discover hidden relations between any drug and any disease: these relations are specific paths among the biomedical entities of the graph, representing possible Modes of Action for any given pharmacological compound. These paths are ranked according to their relevance, exploiting a measure induced by a stochastic process defined on the graph. Here we show, providing real-world examples, how the method successfully retrieves known pathophysiological Mode of Actions and finds new ones by meaningfully selecting and aggregating contributions from known bio-molecular interactions. Applications of this methodology are presented, and prove the efficacy of the method for selecting drugs as treatment options for rare diseases.**

In pharmaceutical research the subject of *drug repurposing* is rapidly raising significant interest. Repurposing means redirection of clinically advanced or marketed products into certain diseases rather than in the initially intended indications. A significant advantage of repurposing drugs is their demonstrated clinical pharmacological efficacy and safety profile. Repurposing is especially interesting in the area of life-threatening Rare or Orphan diseases with high unmet medical need. The hypothesis for drug repurposing is based on the drugs' side effects profiles, indicating interaction with more than one cellular target. These pathway interactions open up the opportunity to exploit existing medicines towards other diseases.

Extensive data sets describing drug effects have been published globally, resulting in a huge amount of information publically available in large on-line collections of bio-medical publications such as PubMed (http://www.ncbi.nlm.nih.gov/pubmed/).

This is an opportunity for literature-based scientific discovery; see [1], [2] and [3]. However, important pieces of information regarding chemical substances, biological processes and pathway interactions are scattered between publications from different communities of scientists, who are not always mutually aware of their findings. In order to generate a working hypothesis from such a body of literature, a researcher would need to read thoroughly all the relevant publications and to pick among them the relevant items of information. Search engines help scientists in this endeavour, but are unable to semantically aggregate information from different sources, leaving all the initiative to researchers; complex relation-focused and graph-like representations (*ontologies*) have been extensively produced and used to fill the gap, since their introduction for the Semantic Web; see [4] and [5]. Yet ontologies need to be man-made and they are difficult to integrate each other and to maintain; see [6].

Here we propose an approach to literature-based research based on the *distributional hypothesis of linguistic theory* (see [7] and [8]) - whose analysis relates the statistical properties of words association to the intrinsic meaning of a concept - and *network theory* (see [9, 10, 11]) - a collection of versatile mathematical tools for representing interrelated concepts and analyse their connections structure.

Main aim of this work is to provide a methodology for creating network knowledge representations, capturing the essential entities occurring in a variety of publications and connecting them into a graph whenever they co-occur in a given sentence. The knowledge graph thus created can then be analysed in order to identify and rank statistically relevant *indirect connections* among prospect medicines and diseases. We show that with a suitable set of concepts, specifically compiled in a dictionary, the linked biochemical entities in the network can be connected along paths that mimic a chain of reasoning and lead to prospect inferences about the mechanism of action of a chemical substance in the pathophysiology of a disease.

In particular, the proposed method consists of two steps: in the first step biomedical papers are collected and submitted to semantic analysis in order to retrieve co-occurrences and compute their similarity; in the second step the knowledge graph, which has been implicitly defined, is analysed to find paths among peptides and rare diseases using techniques derived from the network theory of complex systems.

## INFORMATION RETRIEVAL: TEXT SEARCH

In the field of linguistics it is commonly accepted that the meaning of a word must be inferred by examining its occurrences over large *corpora* of text. Adopting this perspective (see [12] and [13]), one can say that the meaning of a word ultimately depends on the words it mostly goes along with: this is the basis of the so-called "Distributional Hypothesis" introduced by Firth in 1957. The general idea shows that there is a correlation between distributional similarity and meaning similarity, which allows exploiting the former in order to derive the latter. This hypothesis has its shortcomings and intellectual difficulties as an effective definition method. A word can appear often alongside its opposite; if it has many meanings can co-occur with widely different sets of words and its usage may vary with the time (and within different *corpora*). Nonetheless the focus on the connections among words is suggestive, because one can imagine plotting them in a graph in a sort of network of words. The network representation provides new depth to the original linguistic idea because it suggests focusing not only on the relations of a word with those it materially co-occurs with – its *nearest neighbours* in the graph – but with every other word in the network, even those indirectly connected. The meaning of a word is suddenly revealed as something depending on the whole and not on the word itself: in the lingo of complex systems the meaning of a word is an *emergent property* of the language; see [14] and [15].

If this graph of words is transformed into a graph of *concepts* – thereby crossing the boundary between linguistics and semantics –

a network of biomedical entities is obtained, where closeness implies a strong tendency to be connected through predicates. Hopping through this network and drawing a path between any two non-adjacent concepts can be interpreted as suggesting a possible sentence that has never actually been uttered but that can implicitly carry a new and correct idea.

Moreover, it is possible to augment the connection strength between two concepts with the frequency of their co-occurrence, therefore introducing a measure of similarity among concepts, and thus gaining a new tool to gauge the "likeliness" of a relation. This is because highly co-occurring concepts most likely represent tightly bound concepts, if not the same one; the path between two concepts is then also endowed with a similarity weight opening the door to statistical analysis and ranking.

A dictionary, conceived as a list of synonyms, is implemented to transform words into concepts and to work around the disambiguation issue; see [16]. The resolution of synonyms in biology and medicine is well known because there are several names for the same meaning – e.g. the disease Sarcoidosis is also known as Morbus Boeck, or as Besnier-Boeck-Schaumann disease. In the dictionary a list of relevant concepts is defined, comprising peptides, rare diseases and other biological entities such as chemical compounds, proteins, receptors, enzymes, hormones and physiological entities (e.g. cells, organs, tissues, pathways, processes). For each of these concepts a variety of acronyms, synonyms and other identifying phrases – including common typos and language-related different spellings – are collected to be searched for in the papers. Of course different concepts may share some of the identifying expressions. This is the *polysemy* problem (see [17]), i.e. the capacity for a word or a phrase to have multiple meanings. This is a very complex problem in its generality and, although the scientific lingo is quite precise, the methodology provides a disambiguation algorithm, which is a version of the Lesk algorithm (see [17]), to increase precision in concept detection.

Whenever disambiguation fails, we have chosen to keep both the possible concepts: this option reduces precision but maximizes recall – i.e. the quantity of relevant concepts that are retrieved. Many other errors in the detection of co-occurrences arise beyond the ones due to failed disambiguation: a sentence boundary may be misplaced, one of the occurrences may be a false positive or the occurrences may be just part of a list (and therefore not semantically related). It is expected though that as more and more papers are analysed the meaningful co-occurrences will outgrow the spurious ones: in fact "real" co-occurrences are repeated consistently as more and more literature is considered, while spurious ones become statistically insignificant because the same concept is linked randomly to a great number of other concepts. In a figurative manner we may think of a "noise" in the co-occurrence detection that becomes negligible as a large number of papers are considered.

The proposed methodology (Fig. 1) requires that every paper is broken into its constituent sentences and all occurrences of dictionary concepts in them is searched for and registered. This kind of search is carried out either in the whole text or, in principle, in any subset of it. We have chosen the sentence as the text unit to perform the search on to better conform to the distributional hypothesis.

As more and more sentences (and papers) are analysed, more and more co-occurrences are found. Eventually, when all co-occurrences have been retrieved the basic elements are available to build the *knowledge graph*. The nodes of this graph are the entities (concepts) previously defined in the dictionary. A link between two nodes exists if a co-occurrence, involving their corresponding entities, is detected in a sentence. Repeated co-occurrences between the same entities proportionally increase the weight of the link. Therefore this weight is the co-occurrence frequency.

For the purpose of building a knowledge graph dealing with peptides, the related biological processes and the domain of a subset of rare diseases, three million PubMed paper abstracts have been elaborated, out of a total of more than 20 million, with a dictionary of 1606 concepts (127 peptides, 300 diseases and 1179 other biological entities). These are the N vertices of the graph in Fig. 2a.

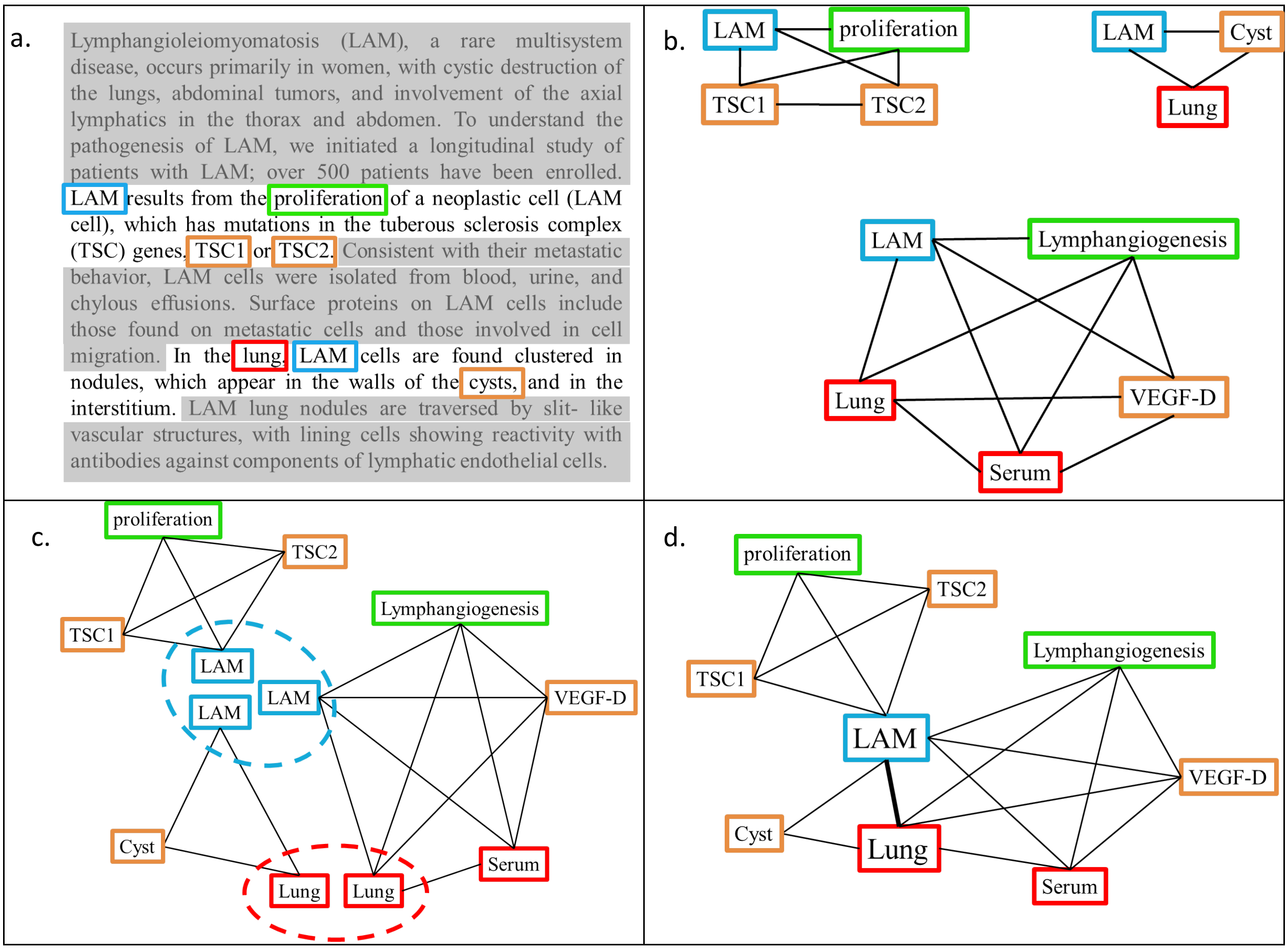

**Figure 1 |Conceptual outline of the knowledge graph building process**. (**a.**) Every document is split into its constituent sentences and each of them is scanned to identify expressions registered on the dictionary. In the figure, two sentences are highlighted and the matching expressions are enclosed in coloured boxes. Every one of these expressions is associated to a concept in the dictionary. (**b.**) The concepts co-occurring in a sentence are connected pairwise. A sentence is therefore abstracted as a complete graph where the occurring concepts are the nodes and a single co-occurrence is a link. The weight of a link is increased if more instances of the same co-occurrence are present. (**c.**) The sentence graphs are then merged in such a way that each node (concept) appears only once in the graph. In the figure it is evident that the «LAM» node (abbreviation for Lymphangioleiomyomatosis – a rare disease) appears in every graph and the «Lung» node in two of them. (**d.**) The result of the merging is a new graph – which is no more complete – where the weight of the link is associated to the frequency of the same co-occurrence.

A single sentence may, of course, contain more than one co-occurrence between concepts, depending on the number of concepts it contains: for a sentence containing $N$ distinct concepts exactly $N(N-1)/2$ co-occurrences are defined, as every concept co-occurs once with every other.

The resulting network is sparse with a small number of links (158,428) compared to the complete graph (12.7%) but, nonetheless, only 30 concepts are not connected to the giant component of the network, thus comprising 1576 nodes (98.13% of the total). The diameter of this network is 4 with an average path length (see [9,11]) of 1.95. It is observed that the graph contains *hubs* interpreted as physiological processes typical of diseases (e.g. inflammation, proliferation, necrosis), immune system-related items (e.g. white blood cells, cytokines) and the major organs – especially the ones dealing with chemical elaboration of drugs (e.g. kidney, liver).

A number of direct connections of "*peptides – diseases"* are present, such as ANGIOTENSIN – SARCOIDOSIS or ANGIOTENSIN – DIABETIC NEPHROPATHY (see Fig. 2b). The relations between those peptides and diseases are already known as we expected on the ground that they appear together in a predicate. Indeed, Angiotensin is known to worsen Sarcoidosis symptoms, while it is of aid in diabetic nephropathy. These features are interpreted as a positive feedback on the meaningfulness of the knowledge graph.

## ANALYSIS OF THE KNOWLEDGE GRAPH

Once the Knowledge Graph is built, we are in the position to analyse it in order to highlight new scientifically analysable relations between a peptide and a rare disease. We search for indirect relations in the network (Fig. 2a) and therefore for a *path* (see [9,11]) between a peptide and rare disease (Fig. 2b). Since all nodes in the network are connected, these paths always exist: the challenge is to rank them (in order to find the most significant ones) and to explore and choose those paths that suggest understandable and yet non-trivial inferences.

Shorter paths must be considered more relevant, as more steps introduce new levels of indirection and magnify the effects of randomness and noise. Yet the paths cannot be too short, because they must be "verbose" enough to suggest a rationale to indicate the biological *Mechanisms of Action* (MoA)*,* i.e. a specific biochemical interaction through which a drug substance produces its pharmacological effect amongst molecular targets like cell receptors, proteins or enzymes; in other words the MoA explains why and how a drug substance works. Specifically, when dealing with peptides, the MoA, that we aim to replicate, is the one where a peptide binds to its specific receptors, thus activating or modulating a physiological process involved in the disease. To achieve such characteristics, we consider specific interactions (links) among nodes, filtering out unwanted information. For instance, a peptide may be connected to any node but since we look for mechanisms of actions, only links in the form of *peptide–cell receptor* are allowed and therefore considered in the graph. Similarly, a receptor can be either involved in a pathway or influence directly a biological process, thus only links in the form *cell receptor–process* or *cell receptor–protein* are allowed.

From the mathematical perspective, co-occurrences define the coefficients of the *similarity matrix A* representing the weighted graph. Through a suitable normalization of *A* we are able to find a probabilistic interpretation for the link weights. Specifically, posing

$$w_{ij} = \frac{a_{ij}}{\sum_k a_{ik}}$$

where $a_{ij}$ is the *i,j* element of the similarity matrix, which is zero if the vertices *i,j* are not directly connected and equal to the edge weight otherwise. Therefore, the components $w_{ij}$ can be interpreted as the conditional probability $p(i|j)$ of finding concept *i* in a sentence containing concept *j*. Since the coefficients of the matrix W are in the range (0,1], we can also introduce a *dissimilarity measure*

$$d_{ij} = -\log(W)$$

which is correctly defined in the range $d_{ij} \in [0, \infty)$ and, oppositely to weights returns larger values for smaller similarities. This distance representation allows immediate application of the available algorithms for computing *shortest paths* (see [11]).

Moreover, shortest paths maximize the sum of link weights, but strong indirect connections between a given peptide and a given disease may arise also from a set of paths which are smaller in weight but that contribute in larger numbers.

The methodology considers all paths connecting the two concepts and uses the abundance and redundancy of these paths, together with their weights, as a measure of the strength of the overall relation between the concepts.

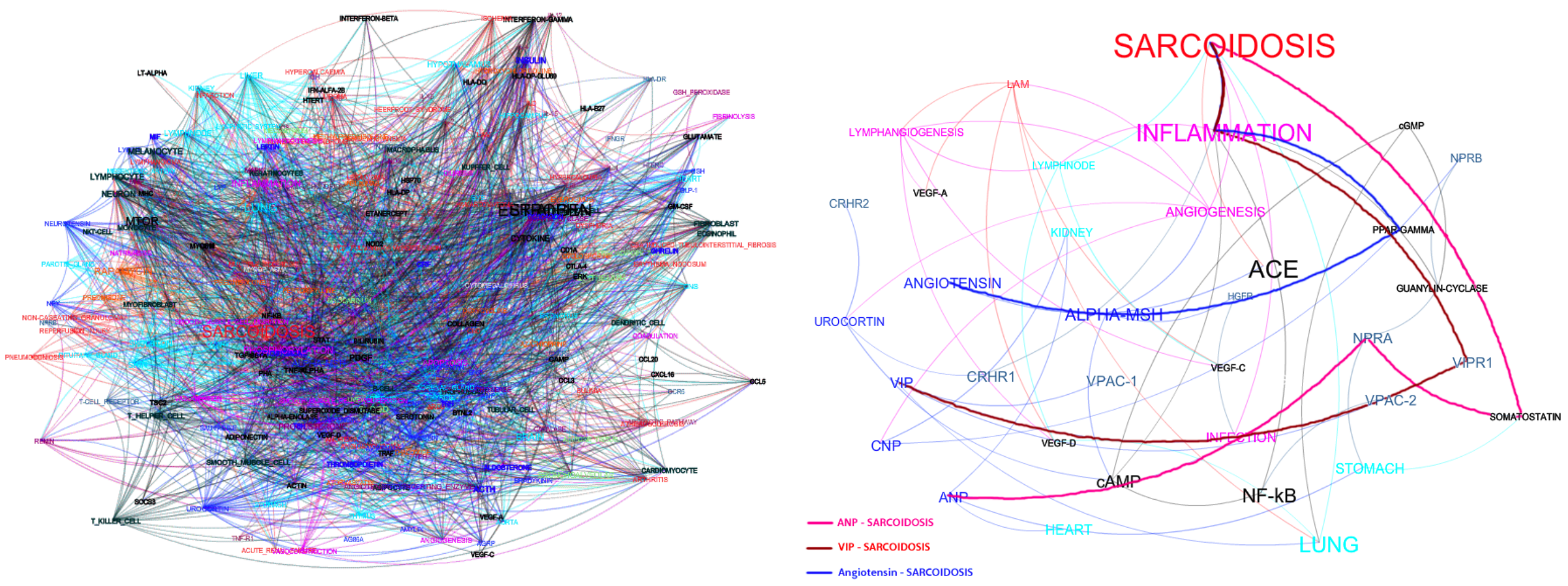

a b

**Figure 2 | Paths identification and selection.** (**a**.) This figure shows a version of the graph – simplified for illustration purposes – built focusing onto 300 concepts and with 200,000 documents. **(b.)** This figure shows three automatically retrieved and meaningful paths, identifying three – out of five – prospect candidate peptides for sarcoidosis. The paths are depicted in a further simplified version of the graph obtained from the first one by filtering out nodes not relevant to the paths.

This can be achieved by measuring the average number of time-steps required to go from one vertex to the other in the network, assuming that a walker is moving at random and that at each discrete time-step it jumps from a vertex to one of its neighbours with a probability which depends on the number of available links and to their weights. This *random walker* produces a distance that is a function of both the length and the abundance of paths.

Intuitively, imagine two nodes connected by one short (one step) path and many longer ones. A random walker trying the route many times will tread the longer paths more often therefore perceiving a "long" distance. Instead, if the end points are connected with a lot of medium-sized paths, the walker will tread those most of the times and thus perceiving a distance shorter than the previous one. A common-world example for conveying this idea: imagine a drunkard trying to go home. He is likely to make many mistakes at the crossroads effectively selecting the next lane at random. He is more likely to get home sooner if many roads converge to his destination rather than if only a short one goes there and the others lead astray.

The random walk distances can be computed by pure algebraic means. The computation is carried out defining a vector, where each component is the likelihood that at a given time a random walker is on a given node. The step-by-step evolution of this vector is a representation of the shifting distribution of these walkers in the nodes in their random wandering.

The probability to walk from vertex *i* to vertex *j* is defined in the random walk theory by the *transfer matrix* P, computed from the similarity matrix *A* with the formula:

$$\mathrm{p_{ij}} = \frac{\mathrm{a_{ij}}}{\sum_k \mathrm{a_{ik}}} \qquad i,j = 1, \dots, N$$

which is exactly the matrix we have previously denoted W. It has been shown (see [18]) that the random walk distances of two nodes *i* and *j* are given by the formula:

$$d(i,j) = \sum_{k=1}^{N} \left[\frac{1}{I - B(j)}\right]_{ik}$$

where *I* is the identity matrix and B is a square matrix identical to P having posed

$$B(j)_{ij} = 0 \quad \forall i\,.$$

The *random walk distance* built this way is non-symmetric, but for our purposes we symmetrise it by taking the average of the two directions.

$$d_s(i,j) = \tfrac{1}{2}[d(i,j) + d(j,i)]$$

This distance defines an implicit ranking measure for each couple of distinct nodes and therefore between any *peptide-disease* couple.

Such a measure can be interpreted as the probability of finding that path, and thus the MoA, within the document base.

**RESULTS**

In this paper we show examples of rationales produced by our methodology with regard to a) the granulomatous disease *Sarcoidosis* and its pulmonary pathology, and b) Imatinib, a targeted-therapy agent against cancer cells, well known for its apoptosis action.

Sarcoidosis is a disease in which abnormal collections of chronic inflammatory cells form as nodules (granulomas) in multiple organs. Sarcoidosis is present at various level of severity in all-ethnic and racial groups and is mainly caused by environmental agents in people with higher genetic sensitivity. The disease is a chronic inflammatory disease that primarily affects the lungs but can affect almost all organs of the body. Sarcoidosis is a complex disease displaying incorrect functionalities within immune cells, cytokines, and antigenic reactions; see [19]. Fig. 3 shows a subgraph of the knowledge network comprising the concepts related with Sarcoidosis.

We were interested in using peptides to treat Sarcoidosis. Therefore a number of rationales have been obtained from a pool of peptides against sarcoid pathologies, and the most relevant findings are listed below, ranked according to the random walk distance:

1. **VIP** – VIPR1 – INFLAMMATION – **SARCOIDOSIS**
2. **α-MSH** – HGFR – INFECTION – **SARCOIDOSIS**
3. **CNP** – NPRB – GUANYLIN_CYCLASE – INFLAMMATION – **SARCOIDOSIS**

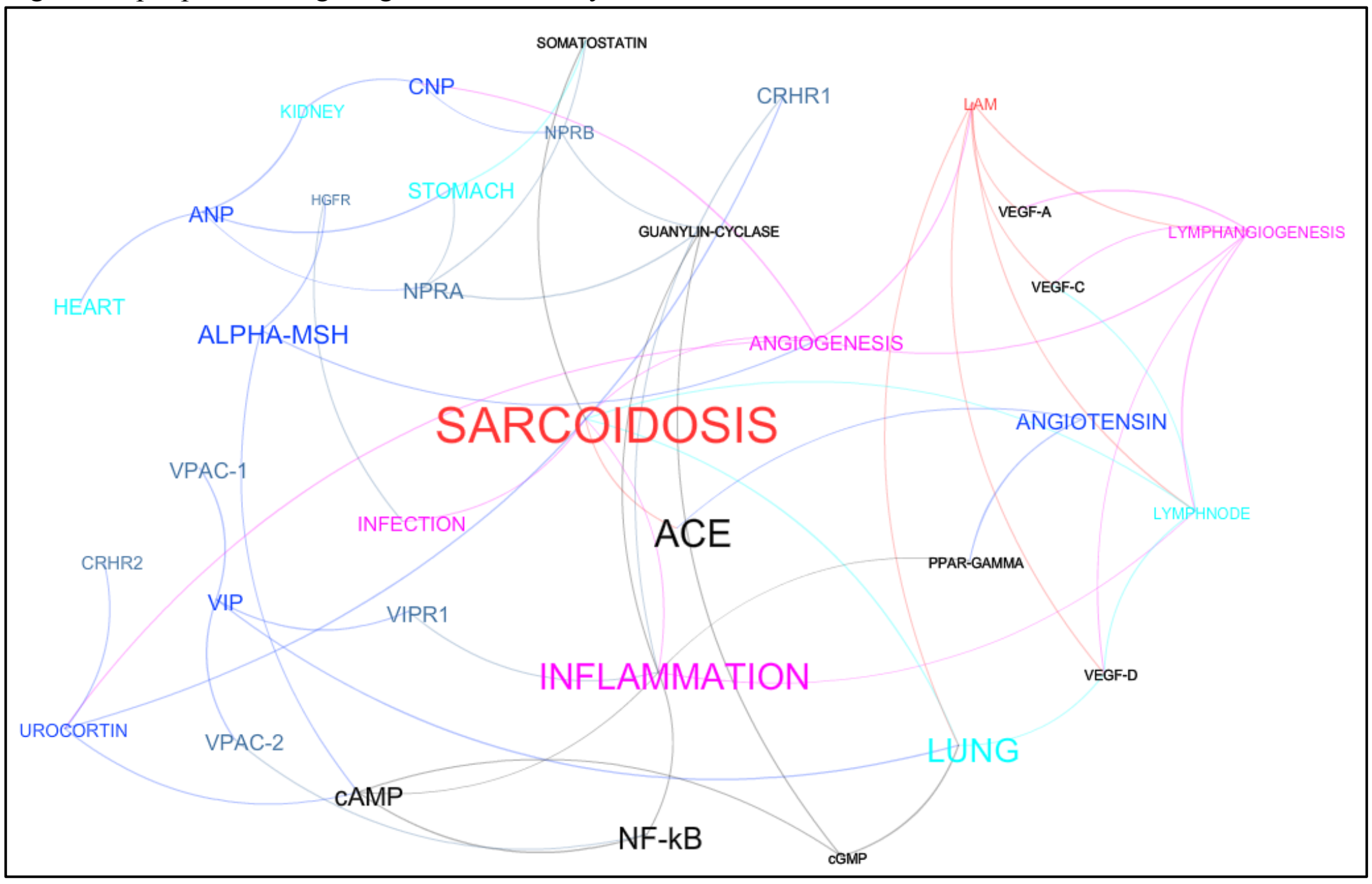

**Figure 3 | The Sarcoidosis knowledge network**. A portion of the knowledge network showing the neighbourhood of Sarcoidosis. The figure is intended as a bird-eye view of the entities the system detected as related with Sarcoidosis

**1. The Match VIP – SARCOIDOSIS**

Vasoactive Intestinal Peptide - VIP (also known as Aviptadil), is an endogenous human peptide. It is predominantly localized in the lungs where it binds specific receptors (VPAC-1, VPAC-2), which transform the signal into an increased production of intracellular cyclic adenosine monophosphate (cyclic AMP or cAMP), as well as into the inhibition of translocation of NF-κB from cytoplasm into the nucleus. This process regulates the production of various cytokines responsible for the inflammatory reaction, such as TNF-α. Hence, VIP is responsible for preventing or attenuating a wide variety of exaggerated pro-inflammatory activities; see [20].

The path in Fig. 4 shows that VIP is affecting the inflammation processes related to Sarcoidosis.

The scientific evidence clearly suggests VIP as a potential treatment option for Sarcoidosis: the system has been able to retrieve the main receptor of VIP and its relevance in the inflammation process.

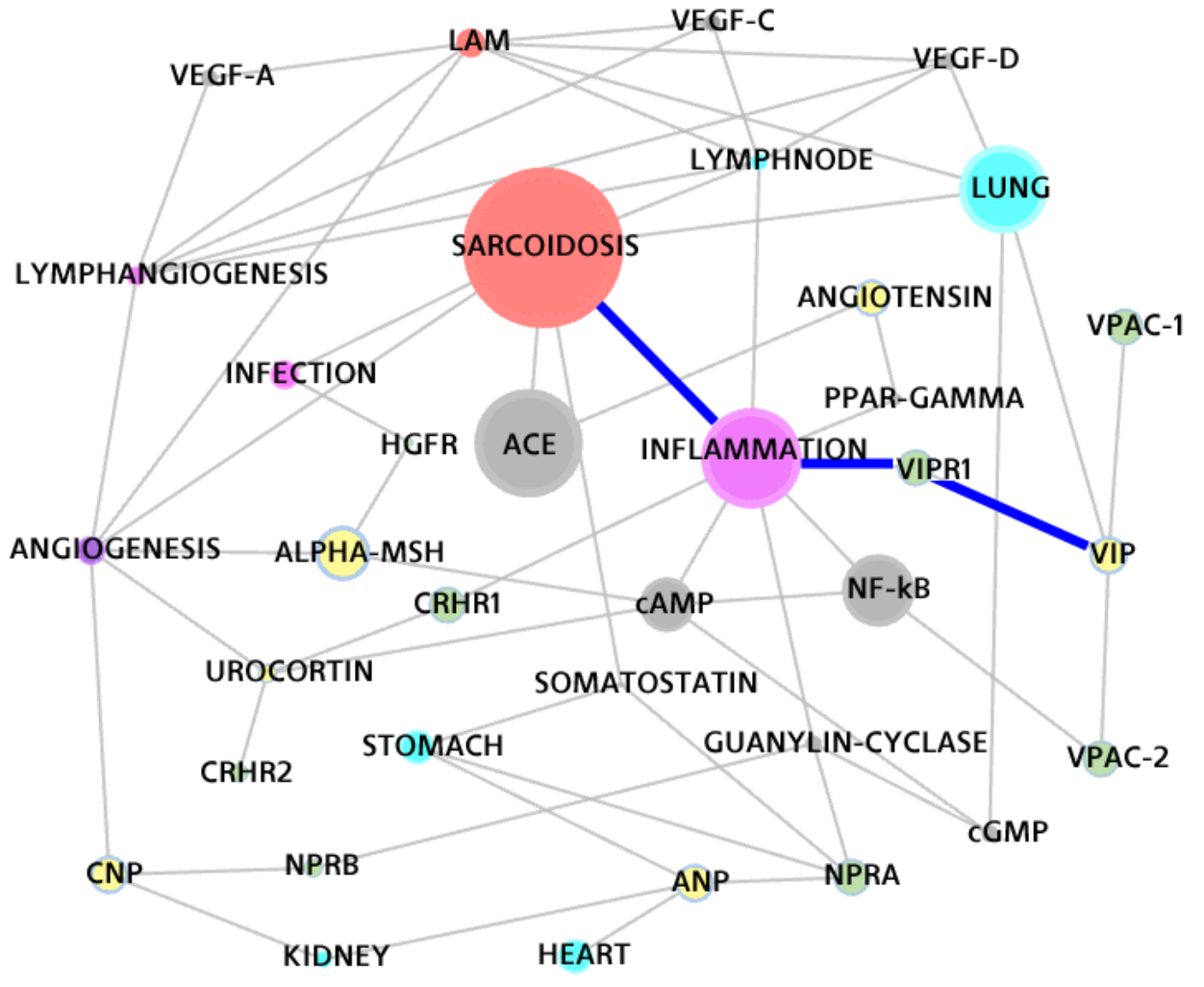

**Figure 4 | The VIP – SARCOIDOSIS path and other closely related concepts.**

## 2. The Match ALPHA-MSH – SARCOIDOSIS

α-Melanocyte Stimulating Hormone (α-MSH) is an endogenous peptide originally described for stimulating melanogenesis, mainly for the pigmentation of the skin. Later it gained roles in feeding behaviour, sexual activity, immune responses, inflammation and fibrosis. Upon binding to its specific cell surface receptors it increases production of cAMP in the target cells and triggers four signalling pathways leading to the disruption of the transcription of several pro-inflammatory mediators genes; see [20].

In addition, α-MSH also regulates the MET proto-oncogene expression in both melanoma cells and in normal human melanocytes. The MET proto-oncogene encodes for the Hepatocyte Growth Factor Receptor (HGFR) that is involved in melanocyte growth and melanoma development; see [21].

There is evidence of interrelation between Epstein-Barr Virus (EBV) infection and MET proto-oncogene expression, and at date several infection agents have been suggested to have an implication as cause of Sarcoidosis.

A role for a transmissible agent is also suggested by the finding of granulomatous inflammation in patients without Sarcoidosis who received heart transplantation from donors who had Sarcoidosis; see [22] and [23].

The system sees both these processes (as apparent from Fig. 5), assigning a better ranking to the second one. α-MSH is another candidate for the treatment of sarcoid pathology due to this double action.

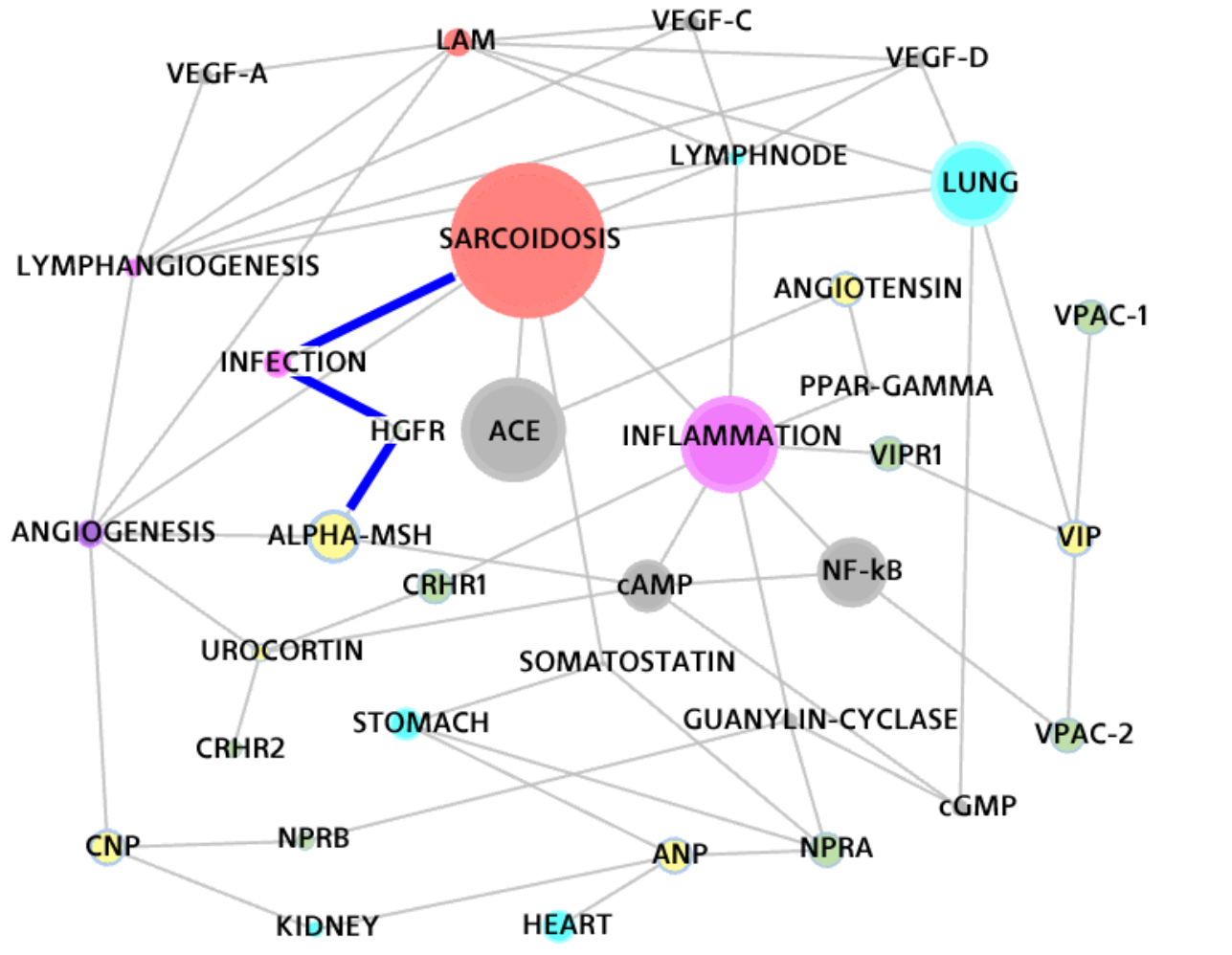

**Figure 5 | The α-MSH – SARCOIDOSIS path and other closely related concepts.**

## 3. The Match CNP – SARCOIDOSIS

CNP (C-type Natriuretic Peptide) is a human peptide, which elicits a number of vascular, renal, and endocrine activities, regulating blood pressure and extracellular fluid volume. When CNP binds to its receptor, NPRB, on the cell surface it activates a cell signalling through a Guanyl cyclase that increases intracellular cGMP level activating specific pathways ultimately modifying cellular functions. cGMP is known for its potent vasodilatory action in pulmonary vessels. Depending on the tissues involved, however, some of its effects are directly opposite to those of cAMP, which is a potent inhibitor of proinflammatory tumor necrosis factor (TNF-α) synthesis; see [24].

The inference subtended by the path in Fig. 7 is sound and correctly traces a biological process. Yet CNP is not considered a treatment option for Sarcoidosis because of its potential negative side effects profile due to its systemic vasodilatory characteristics.

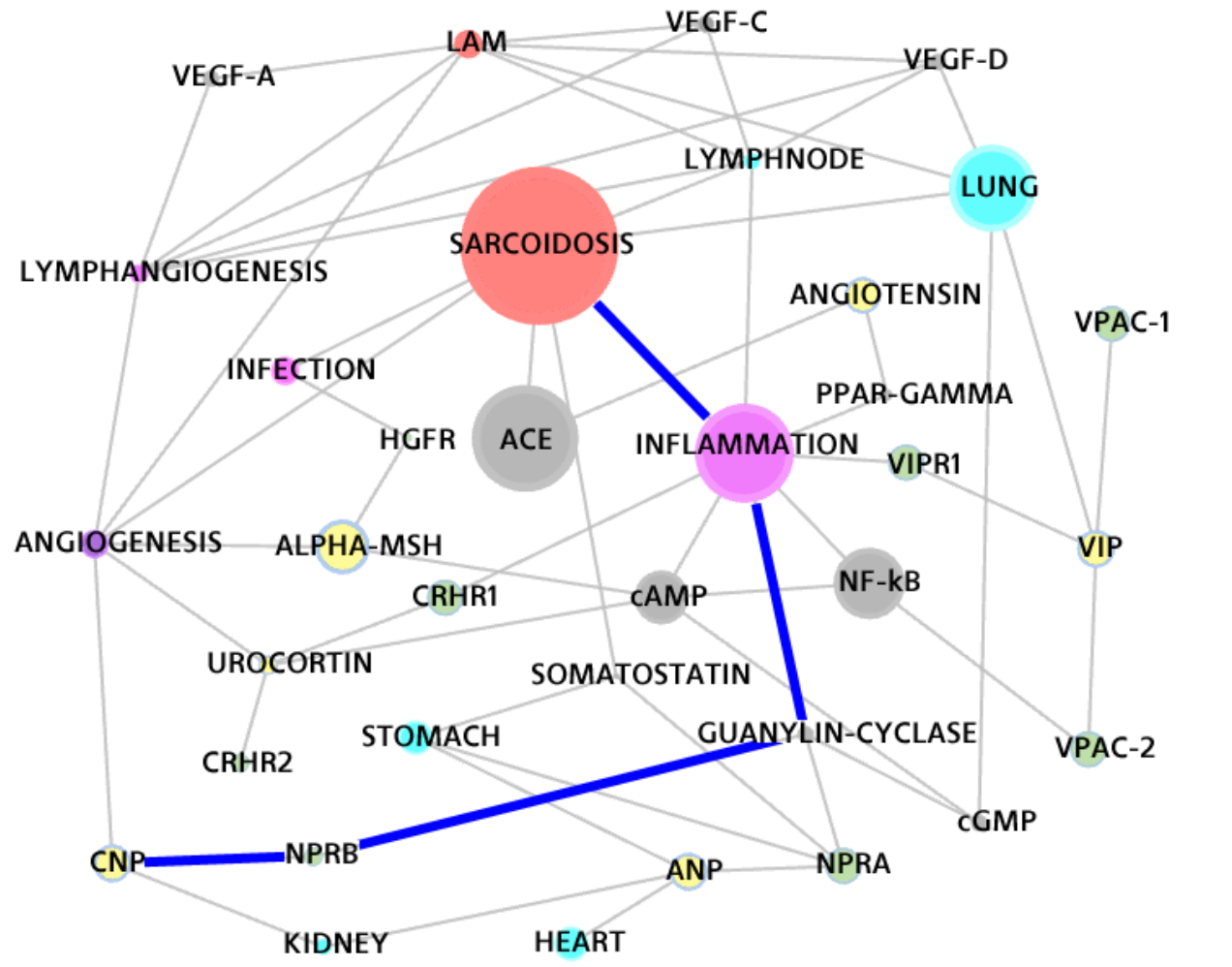

**Figure 6 | The CNP – SARCOIDOSIS path and other closely related concepts.**

## 4. The Match Imatinib – Creutzfeldt-Jakob disease

Imatinib (commercialized under the name GLEEVEC) is a rationally designed pyridylpyrimidine derivative, and a highly potent and selective competitive tyrosine kinase inhibitor, especially effective in the inhibition of kinases c-Abl (Abelson proto-oncogene), c-kit, and PDGF-R (platelet-derived growth factor receptor); see [25] and [26]. These kinases are enzymes involved in cellular signal transduction processes, whose dysregulation may lead to malfunctioning of cells and disease processes, as exemplified in a variety of hyperproliferative disorders and cancers. Imatinib has been regulatory approved for chronic myelogenous leukemia (CML), gastrointestinal stromal tumors (GISTs), aggressive systemic mastocytosis (ASM), hypereosinophilic syndrome (HES), chronic eosinophilic leukemia (CEL), dermatofibrosarcoma protuberans, and Acute Lymphoblastic Leukemia (ALL).

Exploiting our methodology we looked for rationales for the redirection of Imatinib; on the basis of the results of the stochastic measure, the system indicates the neurodegenerative transmissible spongiform encephalopathies – exemplified by the Creutzfeldt-Jakob disease (CJD) – as promising targets for this drug. Transmissible spongiform encephalopathies are caused by the aberrant metabolism of the prion protein (PrP). Prions are seemingly infectious agents without a nucleic acid genome. Prion diseases belong to the group of neurodegenerative diseases acquired by exogenous infection and have a long incubation period followed by a clinical course of progressive dementia, myoclonal ataxia, delirious psychomotor excitement, and neuronal death; see [27].

Moreover, the system selects the path (see Fig. 7) that indicates the kinase c-Abl effect on cell-apoptosis as key MoA for redirecting Imatinib towards CJD. In fact, the c-Abl tyrosine kinase is found to be over-activated in neurodegenerative diseases like Alzheimer's disease and Parkinson's diseases, and overexpression of active c-Abl in adult mouse neurons results in neurodegeneration and neuroinflammation; see [28]. There is clear experimental evidence that activation of c-Abl leads to neuronal cell death and neuronal apoptosis in experimental Creutzfeldt-Jakob disease; see [29]. Imatinib has been shown to prevent c-Abl kinase induced apoptosis in animal models of neurodegeneration; see [30]. Finally, Imatinib was shown to clear prion-infected cells in a time and dose-dependent manner from misfolded infectious protein without influencing the normal biological features of the healthy PrP, and Imatinib activated the lysosomal degradation of pre-existing misfolded PrP; see [31]. This provides a sound rationale for the proposed redirection.

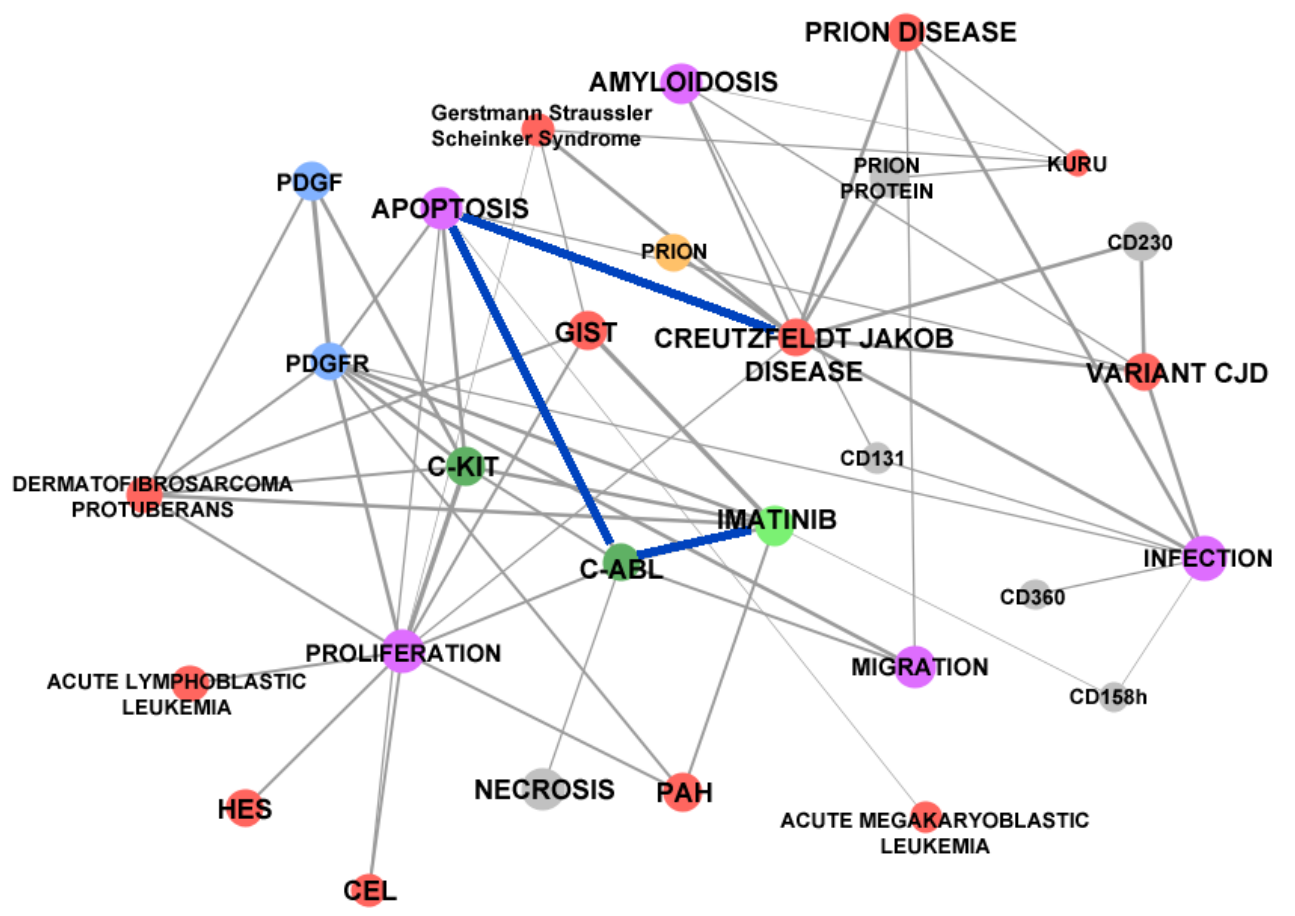

**Figure 7 | Imatinib (GLEEVEC) – Creutzfeldt-Jakob Disease path and other closely related concepts**

The system indicated also Imatinib as a treatment option for pulmonary arterial hypertension (PAH), via its potent inhibitory effect on the PDGF Receptor (PDGF-R). For the indication PAH, the drug is however not approved.

## CONCLUSION

A double-layer methodology is presented, consisting of semantic analysis – leveraging on developments of Computational Linguistics – and graph analysis – exploiting Graph Theory and Stochastic Process Theory tools. This methodology has allowed the screening of more than 3 million abstracts from PubMed-published biomedical papers and the detection of relevant concepts identified by dictionary-defined expressions; concepts have been mapped as nodes of a graph, whose links are defined by co-occurrence of concepts across roughly 30 million of sentences. Specifically, the pathophysiological connections between peptides and diseases have been detected in order to provide inferences for biomedical rationales for drug repurposing.

The proposed methodology provides an effective instrument to detect different MoAs of peptides and drugs; though it may not capture the full-detail of the MoAs, it succeeds in making them recognizable by a short chain of biomedical entities. Moreover, the graph representations of biomedical knowledge seen above produces a sound and meaningful representation of the many interrelated concepts of the biomedical discipline; such methodology successfully allows both the validation of existing rationales and the discovery of new ones, a feat usually left to serendipity and intuition. We have translated the scientific rationales in relevant clinical trial settings into new potential treatment options for the affected patients in Sarcoidosis.

We have looked for experimental evidence of our findings on sarcoid pathologies: we have chosen the peptides VIP and α-MSH for drug repurposing. In an open clinical phase II study, we treated 20 patients with histologically proven Sarcoidosis and active disease with nebulized VIP for 4 weeks. This study is the first to show that VIP has clear, positive, immune-regulatory effects in sarcoid patients without any obvious side effects and without systemic immuno-suppression. VIP should therefore be developed as an attractive therapeutic option for patients with pulmonary Sarcoidosis; see [32]. In parallel, we have initiated a clinical ex-vivo trial to prove α-MSH in a sarcoid pathology. Preliminary data clearly suggest a beneficial outcome of the experimentation, clearly suggesting α-MSH as another potential treatment option for this pathology.

Moreover, the case for Imatinib as a treatment option for the Creutzfeldt-Jakob disease shows how the system is able to produce a sound scientific rationale also for non-peptide drugs and with a mechanism of action quite different from the others, thus proving a much wider applicability.

Results are more noteworthy if the relative slimness of the dictionary is taken into account. Better representations are to be expected defining more detailed and more comprehensive dictionaries. Furthermore, Graph Theory tools provide quite an interesting arsenal of instruments that analyse a complex network of nodes (biological and medical concepts) and highlight hidden inferences across biochemical compounds, clinical data and medical concepts.

As it is apparent from this presentation the specific field of application enters the methodology in the broad selection of the document base and in the definition of the dictionary: the inner mechanism of knowledge representation and analysis is quite independent of it.

We would like to stress that here we have provided only very general characterization of the knowledge network and focused onto very well consolidated tools of analysis. But the field of complex networks is currently under massive development, providing ever more subtle indicators of graph features and related techniques of analysis. We therefore think that our reliance on graph representation poses this method in the best position to exploit this development and may well prove to make it mainstream in the field of information retrieval. Further investigations will concern the use of clustering methods to link groups of other molecules with groups of diseases [33,34].

This methodology can be applied to other fields: it can be extended over broader biomedical research, transcending peptides, to study other chemical compounds and also focusing on diseases or other pathologies other than rare ones. Ultimately, it can be applied to any field of research – even outside natural science – provided that a suitable amount of literature is available and that the main issue be the association of a great number of particular facts and observation that do not yet fit into an already understood and comprehensive scheme.

## ACKNOWLEDGEMENTS

Tiziana Di Matteo and Ruggero Gramatica gratefully acknowledge support by COST TD1210 project.